\documentclass[11pt]{article}  % for review and submission
\usepackage{graphicx}  % needed for figures 
\usepackage{bm}        % for math
\usepackage{amssymb}   % for math
\usepackage{epsfig}

\newcommand \fn {\footnote}

\begin{document}

\title{Energy conditions in Jordan frame}

\author{Saugata Chatterjee${}^1$\fn{schatte8@asu.edu}, Damien A. Easson${}^1$\fn{easson@asu.edu}, and Maulik Parikh${}^{1,2}$\fn{maulik.parikh@asu.edu} \vspace{0.2in}\\
${}^1$Department of Physics, Arizona State University, \\
Tempe, Arizona 85287, USA \vspace{0.1in} \\
${}^2$Beyond Center for Fundamental Concepts in Science, \\
Arizona State University, Tempe, Arizona 85287, USA\\
}

\date{}
\maketitle

%\pacs{98.80.Cq}

\begin{abstract}
The null energy condition, in its usual form, can appear to be violated by transformations in the conformal frame of the metric. We propose a generalization of the form of the null energy condition to Jordan frame, in which matter is non-minimally coupled, which reduces to the familiar form in Einstein frame. Using our version of the null energy condition, we provide a direct proof of the second law of black hole thermodynamics in Jordan frame.
\end{abstract}

%\date{\today}

%%%%%%%%%%%%%%%%%%%%%%%% SECTION %%%%%%%%%%%%%%%%%%%%%%%%%%%%%%
\section{Introduction}
\label{sec:intro}

By themselves, Einstein's equations impose virtually no restrictions on the kinds of spacetimes that are physically permissible. Any symmetric, suitably differentiable metric that satisfies the boundary conditions is a solution to Einstein's equations since, formally, one can simply equate the energy-momentum tensor to the left-hand side of Einstein's equations.

To restrict the solution space, one can impose some physical requirements. In particular, energy conditions are imposed on the types of matter considered. The various energy conditions --- null (NEC), weak (WEC), dominant (DEC), and strong (SEC) --- each express some seemingly reasonable expectation regarding matter, such as that the speed of energy flow be no greater than the speed of light. The energy conditions are inequalities that apply locally, and are asserted to hold everywhere in spacetime. They are generalizations of the notion that local energy density be non-negative and are implemented by requiring various linear combinations of components of the matter energy-momentum tensor to be non-negative. The energy conditions are violated by potentially pathological forms of matter, such as certain instances of tachyons (WEC) and ghosts (NEC), and in turn eliminate many of the arbitrary metrics that would otherwise tautologically satisfy Einstein's equations.

Furthermore, the energy conditions play a critical role in a variety of important theorems in general relativity. They are crucial, for example, to the singularity theorems which indicate that our universe began with a Big Bang singularity; almost all theoretical attempts to evade the Big Bang singularity require the violation of at least some of the energy conditions at some step. Energy conditions are also invoked in the topological censorship theorem, in positive energy theorems, in prohibiting time machines, in the black hole no-hair theorem and, in particular, in the area-law increase for black holes \cite{Bekenstein:1973ur, Bardeen:1973gs, Davies:1987ti}.

Given their importance, it is disturbing that the energy conditions are not derived from any fundamental principles. Indeed, the status of the energy conditions --- why or even whether they hold --- remains unclear \cite{Barcelo:2002bv}. For example it is unclear what happens to the energy conditions in higher derivative theories of gravity; should they be modified, or do the physical laws that depend on them (viz. the laws of black hole thermodynamics) require modification? In $f(R)$ gravity \cite{Jacobson:1995uq} and Brans-Dicke gravity  \cite{Kang:1996rj} no modification is required. But the question is still open for other higher curvature theories of gravity. The NEC is routinely violated in higher derivative theories 
\cite{Nicolis:2008in,Chow:2009fm,Deffayet:2010qz,Hinterbichler:2012yn}, in the presence of extra dimensions \cite{Steinhardt:2008nk}, in ghost condensate models \cite{ArkaniHamed:2003uy}, in string cosmology \cite{Nayeri:2005ck, Brustein:1997cv} and in other cosmological models which evade the big-bang singularity with a cosmic bounce \cite{Khoury:2001wf, Easson:2011zy, Cai:2012va}. The presence of quantum effects and backreaction also tend to introduce ambiguities in the energy conditions \cite{Flanagan:1996gw, Urban:2009yt}.
But even if we restrict ourselves to only lowest-curvature tree-level classical actions, ambiguities appear when non-minimal couplings are introduced \cite{Barcelo:1999hq}. Indeed, even simple conformal frame transformations seem to violate the energy conditions \cite{Visser:1999de}. The reason for this is easy to understand: the energy conditions are conditions on the stress tensor of matter but Weyl transformations (local rescalings of the metric) mix matter and gravity, thereby altering the stress tensor nontrivially.

In this paper, we consider gravity coupled to a scalar field in Einstein frame and show that, after a local rescaling of the metric, the usual form of the null energy condition can appear to be violated. We then propose a new form of the null energy condition, (\ref{eq:NEC-jordan-omega}), that is valid in all rescaled Weyl frames. Our new null energy condition reduces to the usual form in Einstein frame. We then show that this form is consistent with the invariance of the second law of thermodynamics for black holes under changes in conformal frame. We use our modified NEC to supply a direct proof of the second law in Jordan frame.

The result motivates the following conjecture. The energy conditions, unlike the second law, do not appear to be fundamental. We propose that the second law should be taken as given. Then the correct modification of the NEC in a given theory of gravity is that condition on matter that ensures that a classical black hole solution of the theory has an entropy that grows with time.

%%%%%%%%%%%%%%%%%%%%%%%% SECTION %%%%%%%%%%%%%%%%%%%%%%%%%%%%%%
\section{Energy conditions in Jordan frame}
Consider a minimally coupled, Einstein-Hilbert scalar field action with canonically normalized scalar field $\phi$:
\begin{eqnarray}
I =  \int d^D x \sqrt{-g} \left(\frac{M_P^{D-2}}{2} R - \frac{1}{2} (\partial \phi )^2 - V(\phi) \right) \; .  
\end{eqnarray}
Here $g_{ab}$ is the Einstein-frame metric that couples minimally to the metric. The Einstein-frame energy-momentum tensor of the scalar field, $T^E_{ab}$, manifestly satisfies the null energy condition:
\begin{equation}
T_{ab}^E k^a k^b = (k \cdot \partial \phi )^2 \geq 0 \; , \label{eq:NEC-Einstein}
\end{equation}
for any null vector  $k^a$.
Let us consider the modification of the Lagrangian due to a Weyl transformation, $\tilde{g}_{ab} = \Omega^2(x) g_{ab}$.
The existence of a well-defined inverse of the metric requires a nowhere-vanishing conformal factor. In addition, we need $\Omega(x)$ to be related to the previously existing fields in Einstein frame; otherwise, we would have introduced a new degree of freedom. We therefore take $\Omega(x)  = e^{\zeta \phi(x) /\mu}$ with $\mu= M_P^{\frac{D-2}{2}}$ where $\zeta$ is a dimensionless constant, and $\mu$ is a dimensionful constant with the same dimensions as $\phi$. We have chosen the conformal factor to be linear in $\phi$ for simplicity; our results can trivially be extended to a general conformal factor of the form $\exp (f (\phi(x)/\mu))$.

It is important to emphasize here that we have merely redefined field variables from $(g_{ab}(x), \phi(x))$ to $(\tilde{g}_{ab}(x), \phi(x))$. Although the action is typically not invariant under such transformations (unless the metric transformation was induced by a diffeomorphism), the physics is exactly the same \cite{Flanagan:2004bz} provided only that the number of degrees of freedom is unchanged, which is the case here. (Indeed, this is true not only classically, but also quantum-mechanically. In the path integral, the fields are just integration variables. We can redefine these, just as we are free to redefine integration variables in ordinary calculus.) Field redefinitions between the matter and gravitational sectors are used routinely in both string theory and cosmology.
In particular, if the energy conditions are to have physical significance, they too must continue to hold after field redefinition. Let us check whether this is the case.

Using the above choice of conformal factor the action reduces to
\begin{eqnarray}
&& \int \!d^D \!  x \! \sqrt{-\tilde{g}} e^{-\zeta (D-2)\phi/\mu} \! \left(\frac{M_P^{D-2}}{2}  \tilde{R}  - \frac{1}{2} \left( 1  - \zeta^2 (D-2)(D-3) \right) (\partial_a \phi )^2   - e^{-2 \zeta \phi/\mu}  V(\phi)  \right) \; . \quad  \label{sjdf}
\end{eqnarray}
We will refer to this action as the Jordan-frame action because gravity is non-minimally coupled, via the term in which the Ricci scalar is directly coupled to matter. The various pieces of the Lagrangian no longer split cleanly into a gravity Lagrangian and a matter Lagrangian, and hence it is no longer sensible to define the Jordan-frame stress tensor, $T^J_{ab}$, naively, as the variation of the matter action:
\begin{equation}
T^J_{ab} \neq -\frac{2}{\sqrt{-\tilde{g}}} \frac{\delta I_{\rm matter}}{\delta \tilde{g}^{ab}} \; .
\end{equation}
However, The equation of motion for $\tilde{g}_{ab}$ can be rewritten so as to get the Einstein tensor on the left-hand side. Then whatever is on the right-hand side is covariantly conserved as a result of the Bianchi identity. Thus we can simply define the Jordan frame stress tensor as the quantity proportional to the Einstein tensor in the gravitational equations:
\begin{equation}
\tilde{G}_{ab} \equiv \frac{1}{M_P^{D-2}} T_{ab}^J \; . \label{eq:EOM-jordan}
\end{equation}
Explicitly, the stress tensor in Jordan frame is
\begin{eqnarray}
  &&T_{ab}^J =  (1 + \zeta^2 (D-2))(\partial_a \phi ) (\partial_b \phi ) - \tilde{g}_{ab}\left(\frac{1}{2} \left(1 + \zeta^2 (D-1)(D-2) \right)  (\partial_a \phi )^2 + e^{-2 \zeta \phi/\mu} V( \phi)\right) \nonumber \\
 && \quad \quad \quad \quad  + \zeta (D-2) M_P^{\frac{D-2}{2}} \left( \tilde{g}_{ab} \tilde{\nabla}^2 - \tilde{\nabla}_b \tilde{\nabla}_a \right) \phi \; ,  \label{eq:Tab-jordan} 
\end{eqnarray}
which is covariantly conserved:
\begin{equation}
\tilde{\nabla}^a T_{ab}^J = 0 \; .
\end{equation}

%%%%%%%%%%%%%%%%%%%%%%%% subsection %%%%%%%%%%%%%%%%%%%%%%%%%%%%%%

\subsection*{NEC on the stress tensor in Jordan frame}

The covariantly-conserved stress tensor, (\ref{eq:Tab-jordan}), can still violate the null energy condition in its Einstein-frame form \cite{Visser:1999de}. To see this, contract $T^J_{ab}$ with null vectors:
\begin{equation}
T_{ab}^J k^a k^b = \left(  1 +  \zeta^2 (D-2) \right) (k^a \partial_a \phi )^2  - \zeta (D-2) M_P^{\frac{D-2}{2}} ~ k^a k^b  \tilde{{\nabla}}_a \tilde{{\nabla}}_b \phi \; .
\end{equation}
Notice that the positivity of the last term is ambiguous. 
If this were the correct form of the NEC, the set of physical solutions $(g_{ab}, \phi)$ in Einstein frame --- namely, those that obey the Einstein frame NEC --- would not map to the set of physical solutions $(\tilde{g}_{ab}, \phi)$ in Jordan frame --- those that satisfy $T^J_{ab} k^a k^b \geq 0$. But the set of physical solutions should not change under field redefinitions. It must be, then, that this naive version of the NEC is incorrect in Jordan frame.

We propose that the correct null energy condition in Jordan frame is
\begin{equation}
\left [ T_{ab}^J  + \zeta (D-2) M_P^{\frac{D-2}{2}}  \tilde{{\nabla}}_a \tilde{{\nabla}}_b \phi \right ] k^a k^b \geq 0 \; . \label{eq:NEC-jordan}
\end{equation}
Once this condition is imposed on the matter fields in the Jordan frame the ambiguity is resolved and we have (solution set in Jordan frame) $\equiv$ (solution set in Einstein frame). To see this explicitly, consider our energy-momentum tensor. Using the expression (\ref{eq:Tab-jordan}) in the relation (\ref{eq:NEC-jordan}), we find
\begin{equation}
\left ( T_{ab}^J +  \zeta (D-2)  M_P^{\frac{D-2}{2}} \tilde{{\nabla}}_a \tilde{{\nabla}}_b \phi \right ) k^a k^b = \left(  1 +  \zeta^2 (D-2) \right) T^E_{ab} k^a k^b \; .
\end{equation}

Hence, (modified NEC obeyed in Jordan frame) $\Leftrightarrow$ (usual NEC obeyed in Einstein frame). Our expression for the Jordan frame null energy condition can be generalized to arbitrary non-minimal couplings of matter sources to the Ricci scalar. The generalized version is
\begin{equation}
\left [T^J_{ab}  + (D-2) M_P^{D-2}  \tilde{\nabla}_a \tilde{\nabla}_b  \ln \Omega \right ] k^a k^b \geq 0 \; .\label{eq:NEC-jordan-omega}
\end{equation}
This inequality reduces to the usual Einstein frame NEC, (\ref{eq:NEC-Einstein}), in case of minimal coupling ($\Omega=1$) and to the modified NEC, (\ref{eq:NEC-jordan}), when $\Omega = \exp(\zeta \phi(x)/\mu)$.

To summarize, we propose a new form of the null energy condition, (\ref{eq:NEC-jordan-omega}), that applies to non-minimally coupled scalar fields and whose solution set is the same as that of the usual Einstein frame null energy condition. This is as it should be, since the two frames are related by a field redefinition.

%%%%%%%%%%%%%%%%%% SECTION %%%%%%%%%%%%%%%%%%%%%%%%%

\section{Entropy increase in Jordan frame} \label{sec:entropy-increase}

Our proposal for the modified null energy condition in Jordan frame was somewhat ad hoc. It happened to work for non-minimally coupled scalar fields in conformally transformed actions. We will now motivate the prescription by a robust physical principle. If we regard black hole entropy as counting the number of degrees of freedom via the dimensions of its Hilbert space, or alternatively, the number of possible initial configurations from which the black hole could be formed, then the entropy is clearly conformally invariant.
If the entropy increases in one frame as a result of imposition of the NEC then it should also increase in the conformal frame without any extra requirement, viz. the modified NEC of the Jordan frame. In this section, we will prove that our modified null energy condition in Jordan frame indeed guarantees that, classically, black hole entropy never decreases.

Consider a stationary black hole solution to the Einstein equation in $D$ spacetime dimensions. 
The entropy of black hole solutions of the Einstein-Hilbert action is just proportional to its ``area" \cite{Bekenstein:1973ur}, by which we mean a (D-2)--dimensional spacelike section of the horizon. But in dealing with spacetimes which are solutions to non-Einstein-Hilbert actions, such as the one in Jordan frame with non-minimal scalar coupling, a notion of entropy is absent. The Wald prescription of entropy \cite{wald-iyer} identifies a conserved charge with the entropy of the horizon for stationary solutions to these non-Einsteinian theories of gravity. The correct entropy to use for dynamical horizons is Jacobson-Myers entropy \cite{Jacobson:1993xs}. But since  the metric has rescaled, we can obtain the entropy just by applying a field redefinition to the Bekentein-Hawking entropy \cite{Jacobson:1993vj}. The black hole entropy in Jordan frame is not simply proportional to the area. Rather it is
\begin{equation}
S = \frac{1}{4 G_D} \int d^{D-2} x \sqrt{\tilde{\gamma}} \, \Omega^{-(D-2)} \; .
\end{equation}
The increase of entropy in the Jordan frame has been studied before from the perspective of $f(R)$ theories \cite{Jacobson:1995uq} and the second law was proved in Jordan frame in \cite{Ford:2000xg} using the Einstein-frame NEC. We provide a proof directly in Jordan frame using our modified Jordan-frame NEC.

To see how the rate of change of black entropy in the Jordan frame depends on the Jordan frame NEC we first find the expression for the change in the black hole entropy.
\begin{eqnarray}
 \frac{d S}{d \tilde \lambda} & = &\frac{1}{4 G_D} \int d^{D-2} x \sqrt{\tilde \gamma} \, \Omega^{-(D-2)} \left(  \tilde \theta - (D-2) \frac{d \ln \Omega}{d \tilde \lambda} \right) \nonumber \\
  & \equiv  &\frac{1}{4 G_D} \int d^{D-2} x \sqrt{\tilde \gamma} \, \Omega^{-(D-2)} \Theta \; , \label{2ndlaw}
\end{eqnarray}
where we have defined
\begin{eqnarray}
\Theta(\tilde \lambda) = \tilde \theta - (D-2) \frac{d \ln \Omega}{d \tilde \lambda} \; . \label{Theta}
\end{eqnarray}
Here $\tilde \theta$ is the expansion scalar for the null generator $\tilde k^a$ of the horizon  in the Jordan frame:
\begin{equation}
\tilde \theta =  \tilde{\nabla}_a \tilde k^a = \frac{d (\ln \sqrt{\tilde \gamma})}{d \tilde \lambda} \; .
\end{equation}
Black hole entropy would not decrease if $\Theta$ were non-negative, as evident from (\ref{2ndlaw}). We will now give a direct proof of the second law of thermodynamics for scalar-tensor theories in Jordan frame. As in the proof that the surface area of black holes in Einstein gravity always increases \cite{Bardeen:1973gs}, we will show that, if $\Theta < 0$, then a caustic necessarily forms.

Notice that the vector $\tilde k^a = (d/d \tilde \lambda)^a$ is related to $k^a = (d/d \lambda)^a$. This is easiest to see for a normalized timelike velocity vector, $u^a = (d/d \tau)^a$. Since $d \tau^2 = - g_{ab} dx^a dx^b$, rescaling the metric causes $\tau$ to scale: $\tilde \tau = \Omega \tau$. Then $\frac{d \tilde \tau}{d \tau} = \Omega$ and hence $\tilde{u}^a \equiv (d/d \tilde \tau)^a = \frac{d \tau}{d \tilde \tau} (d/d \tau)^a = (1/\Omega) u^a$. Similarly, we have
\begin{equation}
\tilde k^a \equiv \left ( \frac{d}{d \tilde \lambda} \right )^{\! a} = \frac{d \lambda}{d \tilde \lambda}  \left ( \frac{d}{d \lambda} \right )^{\! a} = \frac{1}{\Omega} k^a \; .
\end{equation}
Because $\Omega(x)$ is a function of space and time, $\tilde k^a$ is not in general affinely parameterized. Thus
\begin{equation}
\tilde k^b \nabla_b \tilde k^a = \kappa \tilde k^a \; ,
\end{equation}
where 
\begin{equation}
\kappa =  \frac{d}{d \tilde \lambda} \ln \Omega \; . \label{kappa}
\end{equation}
The corresponding Raychaudhuri equation for $\tilde \theta$ is 
\begin{equation}
\frac{d \tilde \theta}{d \tilde \lambda} = \kappa \tilde \theta - \left(  \frac{\tilde \theta^2}{D-2} +  \tilde \sigma_{ab} \tilde \sigma^{ab} + \tilde \omega_{ab} \tilde \omega^{ab} + \tilde R_{ab} \tilde k^a \tilde k^b  \right) \; . \label{eq:omega-inc}
\end{equation}
The presence of the $\kappa \tilde \theta$ term on the right is a sign that $\tilde \lambda$ is not an affine parameter. Hypersurface orthogonality of the null generators $\tilde k^a$ of the event horizon causes the rotation ($\tilde w=0$) to vanish by the Frobenius theorem. 
 We can now use the equation of motion in the Jordan frame and replace the Ricci tensor with the Jordan frame stress tensor:
\begin{equation}
\tilde R_{ab} \tilde k^a \tilde k^b = \frac{1}{M_P^{D-2}} T_{ab}^J  \tilde k^a \tilde k^b \; .
\end{equation}
Taking the derivative of (\ref{Theta}) and substituting (\ref{eq:omega-inc}) then gives
\begin{eqnarray}
\frac{d \Theta}{d \tilde \lambda} &=& \frac{d \tilde \theta}{d \tilde \lambda} - (D-2) \frac{d^2}{d \tilde \lambda^2} \ln \Omega \nonumber \\
&=&  \kappa \tilde \theta - \left(  \frac{\tilde \theta^2 }{D-2} + \tilde \sigma_{ab}^2 + \tilde R_{ab} \tilde k^a \tilde k^b  \right) - (D-2) \frac{d^2}{d \tilde \lambda^2} \ln \Omega \nonumber \\
&=&  - \frac{\Theta^2}{D-2}  - \kappa \Theta - (D-2) \kappa^2  - \tilde \sigma_{ab}^2 - \frac{\tilde k^a \tilde k^b }{M_P^{D-2}} \left( T^J_{ab} + (D-2) M_P^{D-2} \tilde \nabla_a \tilde \nabla_b \ln \Omega \right) . \nonumber \\
&& \label{dTheta}
\end{eqnarray}
The last line follows from the relation between $\kappa$ and $\Omega$ above (\ref{kappa}). Note the presence of the terms in square brackets: this is proportional to precisely the expression that appears in our modified null energy condition in Jordan frame, (\ref{eq:NEC-jordan-omega}). When that is obeyed we have
\begin{equation}
\frac{d \Theta}{d \tilde \lambda} \leq - \frac{\Theta^2}{D-2}  - \kappa \Theta - (D-2) \kappa^2 \; . \label{analyzethis}
\end{equation}
We will now analyze this equation carefully in order to prove that only solutions with $\Theta > 0$ do not have caustics. Cosmic censorship --- which in this context means the prohibition of caustics --- eliminates all solutions with $\Theta < 0$ and hence, by (\ref{2ndlaw}), the second law of black hole thermodynamics holds in Jordan frame.

Suppose, then, that at some parameter $\tilde \lambda_0$, a pencil of horizon null generators has $\Theta_0 < 0$. For a sufficiently thin  pencil, the surface gravity is effectively constant over spacelike sections of the pencil. Therefore, we can regard $\kappa$ as  a function  of  $\tilde \lambda$ only. First, consider $\kappa \leq 0$. But then every term on the right-hand side of (\ref{analyzethis}) is nonpositive for $\Theta < 0$. Hence $\frac{d \Theta}{d \tilde \lambda} \leq 0$. In fact, $\frac{d \Theta}{d \tilde \lambda} \leq - \frac{\Theta^2}{D-2}$ whose solution is $\Theta(\tilde \lambda) \leq \Theta_0/(1 + \frac{\tilde \lambda - \tilde \lambda_0}{D-2} \Theta_0 )$. For all negative values of $\Theta_0$, $\Theta(\tilde \lambda)$ diverges at some finite $\tilde \lambda$, resulting in a caustic. Hence, for $\kappa \leq 0$, all solutions with $\Theta < 0$ lead to caustics.

Next, consider $\kappa > 0$. In this case, the term $-\kappa \Theta$ in (\ref{analyzethis}) is positive for $\Theta < 0$. However, the three terms on the right-hand side together are always negative. To see this, consider the right-hand side of (\ref{analyzethis}) as a quadratic polynomial in $\Theta$; this quadratic has no real roots for $\kappa > 0$. Hence again $\frac{d \Theta}{d \tilde \lambda} \leq 0$; $\Theta$ is a monotonically decreasing function of $\tilde \lambda$. However, to prove that this inevitably results in a caustic is more subtle because the positivity of the $-\kappa \Theta$ term does not permit us to write $\frac{d \Theta}{d \tilde \lambda} \leq - \frac{\Theta^2}{D-2}$.

A monotonically decreasing negative function $\Theta(\tilde \lambda)$ can have three different asymptotic possibilities. Possibility 1 is that $\Theta$ asymptotically and monotonically approaches some finite negative value, $\Theta_{\rm min}$ i.e. $\lim_{\tilde \lambda \to \infty} \Theta (\tilde \lambda) = \Theta_{\rm min}$. Possibility 2 is that $\Theta$ is unbounded from below but reaches negative infinity only in the infinite future i.e. $\lim_{\tilde \lambda \to \infty} \Theta (\tilde \lambda) = - \infty$. This is not a caustic because $\Theta$ is finite at all finite values of $\tilde \lambda$. Possibility 3 is that $\Theta$ diverges at some finite $\tilde \lambda_c$: $\lim_{\tilde \lambda \to \tilde \lambda_c} \Theta (\tilde \lambda) = - \infty$. This corresponds to a caustic. These three possibilities are illustrated schematically by the curves in Figure 1. We will now show that $\kappa > 0$ and $\Theta < 0$ always gives rise to possibility 3.

\begin{figure}[hbtp]
 \centering
  \epsfysize=7 cm
  \epsfbox{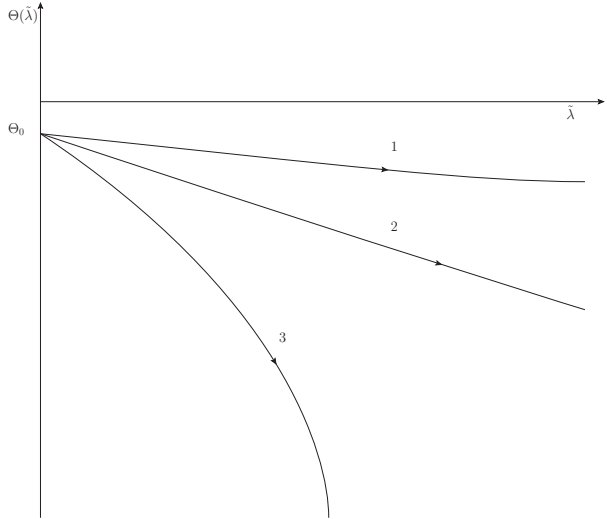}
 \centering
 \caption{Possible curves for a negative monotonically decreasing function $\Theta (\tilde \lambda)$.}
 \label{fig:geometry}
 \end{figure}
 
First we rule out possibility 1; $\Theta(\tilde \lambda)$ does not asymptotically approach a finite value. For suppose that were true. Then for large values of $\tilde \lambda$, we would have $\Theta \approx \Theta_{\rm min}$ and $\frac{d \Theta}{d \tilde \lambda} \approx 0$. But, regarding the right-hand side of (\ref{analyzethis}) as a quadratic in $\kappa$, we see that there are no real solutions for $\kappa$ when $\Theta = \Theta_{\rm min}$ and $\frac{d \Theta}{d \tilde \lambda} = 0$. Hence, possibility 1 is eliminated and $\Theta(\tilde \lambda)$ is therefore unbounded from below.

Since $\Theta$ is unbounded from below, consider a very large (negative) value of $\Theta$. Focus on an infinitesimal interval of $\tilde \lambda$. In that interval, $\kappa(\tilde \lambda)$ can be regarded effectively as a constant. Then we can integrate (\ref{analyzethis}) to obtain
\begin{equation}
\Theta(\tilde \lambda) = \frac{\frac{\sqrt{3}}{2} \Theta_0 - \left( \Theta_0 + \frac{7}{4} (D-2) \kappa \right) \tan \left( \frac{\sqrt{3}}{2} \kappa (\tilde \lambda - \tilde \lambda_0)\right)}{\frac{\sqrt{3}}{2}  + \left( \frac{\Theta_0}{\kappa (D-2)} + 1 \right) \tan \left( \frac{\sqrt{3}}{2} \kappa (\tilde \lambda - \tilde\lambda_0)\right)} \; .
\end{equation}
Scrutiny of this reveals that the denominator vanishes for certain values of $\tilde \lambda$:
\begin{equation}
\tilde \lambda - \tilde \lambda_0 \approx - \frac{D-2}{\Theta_0} \; .
\end{equation}
Hence if $\Theta_0$ is large and negative, $\Theta$ becomes divergently negative in finite time: a caustic.

We have proven that, whether $\kappa$ is positive or negative, we always find a caustic in finite parameter $\tilde \lambda$ whenever $\Theta_0 < 0$. Cosmic censorship bans these solutions leaving only those with $\Theta \geq 0$. This in turn implies that black hole entropy must be nondecreasing in Jordan frame. Our proof relied crucially on (\ref{analyzethis}), which follows from (\ref{dTheta}) only when our modification to the Jordan-frame null energy condition is satisfied. A quick check on our result comes from the observation that $\Theta = \theta_E/\Omega$, where $\theta_E$ is the expansion in Einstein frame. Then, when the usual Einstein-frame null energy condition is satisfied, $\theta_E$ is positive. This in turn means that $\Theta$ must be positive and that the Jordan-frame entropy also increases. Here we have proven that fact directly in Jordan frame without relying on a correspondence with Einstein frame.

\subsection*{Discussion}
By using a field redefinition, we have seen that the form of the NEC is modified in Jordan frame. Similar rewritings lead to modified versions of the other energy conditions, whose forms are not particularly illuminating. However, field redefinitions from Einstein frame do not exist for generic theories, such as most higher-derivative theories. The question naturally arises as to what the appropriate generalization of the NEC is for such theories. A possible clue is to be found in (\ref{dTheta}). Given only that equation plus the requirement that the second law hold, one could in fact have inferred the modified NEC directly invoking neither a field redefinition nor even the original Einstein-frame NEC. This is because all terms on the right-hand side of (\ref{dTheta}) need to be negative in order to guarantee the validity of the second law. This is a very interesting observation because, rather than using the modified NEC to prove the second law, we would in this approach take the second law as a given and derive the appropriate condition for matter. This is appealing because the NEC -- unlike the second law -- does not seem to rest on any fundamental principles of physics. In the same spirit as Jacobson's Einstein equation of state paper \cite{Jacobson:1995ab} (in which thermodynamic laws are taken as axioms rather than as statements to be proved), one should perhaps begin, not end, with the second law. This approach would generalize to other gravitational theories, provided one had the correct formulation of entropy, valid in non-stationary situations.

\bigskip
\noindent
{\bf Acknowledgments}

\noindent
It is a pleasure to thank Paul Davies and Jan Pieter van der Schaar for helpful discussions. This work is supported in part by DOE grant DE-SC0008016.

%%%%%%%%%%%%%%%%%%%%%%%% bibliography %%%%%%%%%%%%%%%%%%%%%%%%%%%%%%
\bibliographystyle{plain}

%%%%%%%%%%%%%%%%%%%%%%%% THE END %%%%%%%%%%%%%%%%%%%%%%%%%%%%%%
\end{document}